\documentclass{article}
\usepackage{graphicx,amsmath,epsfig}
\newcommand{\be}{\begin{equation}}
\newcommand{\ee}{\end{equation}}
\newcommand{\bea}{\begin{eqnarray}}
\newcommand{\eea}{\end{eqnarray}}

\newcommand{\T}{\mbox{\rm T}}

\newcommand{\ud}{d }

\begin{document}

\title{Existence of spinning solitons in field theory\footnote{Talk given 
at the Seventh Hungarian Relativity Workshop, 10--15 August, 2003,
Sarospatak, Hungary. }}

\author{\it Mikhail S. Volkov\\
\rm Laboratoire de Math\'ematiques et Physique Th\'eorique, \\ 
Universit\'e de Tours, Parc de Grandmont, 37200 Tours,
FRANCE \\
E-mail: volkov@phys.univ-tours.fr}

\maketitle

\abstract{
The first results, both positive and negative,
recently obtained in the area of constructing 
stationary spinning solitons in flat 
Minkowski space in 3+1 dimensions are discussed. 
}

\section{Introduction}  \label{intro}

It is well known that in General Relativity there exist classical solutions
describing spinning objects with finite energy -- Kerr-Newman  
black holes. 
They are uniquely characterized by their mass, electric
charge, and angular momentum $J$.  Static and
spherically symmetric solutions with $J=0$ are relatively easy
to obtain, 
and historically they had been found first, while it took then 
more than 40 years
until their stationary generalizations with $J\neq 0$ were constructed.

If one descends from curved space to flat Minkowski space, one finds 
there non-gravitational field 
theories, as for example the 
Yang-Mills theory,  described at the classical level 
by non-linear partial differential equations. 
In some cases these equations admit solutions 
describing localized, globally regular particle-like objects 
with finite energy -- solitons. 
There exist various types of solitons in field-theoretic models,
such as, for example, 
monopoles \cite{mon} (they will be briefly reviewed below),
 dyons \cite{dyon}, vortices \cite{vort}, 
sphalerons \cite{sphal}, 
Skyrmeons \cite{Skyrm}, knots \cite{knot}, Q-balls \cite{Q}, etc.   
In most cases explicitly known soliton solutions are static and spherically
symmetric. It is then natural to wonder whether one can 
find for them stationary, spinning generalizations with $J\neq 0$,
 similar to what has been done for black holes. More generally, one can ask

\vspace{1 mm}

{\sl Do static solitons
admit stationary, spinning generalizations~?
} 

\vspace{1 mm}

\noindent
It is quite natural to conjecture 
 that the answer is positive. 
However, until very recently this intuitive conjecture had neither been 
supported by 
any explicit examples of spinning solitons, nor had it been frustrated 
by any no-go conclusions. 

To be precise, by {\sl spinning} solitons are 
meant here  $J\neq 0$ solutions in the {\sl one-soliton} sector. They describe
rotational, spinning excitations of an individual, isolated object. On the other hand, 
one can also consider rotating solutions outside the one-soliton sector. 
These would rather describe  relative {\sl orbital} motions in many-soliton systems,
such as, for example,  a pair of soliton and antisoliton rotating around 
their common center of mass.
Another example of systems which could be naturally classified as orbiting are
rotating vortex loops -- vortons. 
Solutions describing such orbital rotations are actually known (see \cite{VW02}
for a discussion) and will not be considered here.

In what follows I shall present the first results on the existence of 
spinning solitons
obtained recently in our work  with Erik W\"ohnert \cite{VW02},
\cite{VW03}. 
They  are two-fold, both positive and negative.
The positive statement is:

\vspace{1 mm}

\noindent 
$\bullet$ {\sl Solitons in theories with rigid symmetries can have spinning
excitations.} 

\vspace{1 mm}

\noindent
This is supported by an explicit construction of spinning
solutions. Next, however, comes a  no-go result: 

\vspace{1 mm}

\noindent
$\bullet$ {\sl None of the known solitons in gauge field theories 
with gauge group SU(2) have spinning generalizations within the 
axially symmetric sector.} 

\vspace{1 mm}

\noindent
This fact  
is quite surprising, as it rules out a large class of
spinning solitons, in particular, 
monopoles, dyons, sphalerons, and vortices.  

\section{Explicit example of spinning solitons}

Let us consider a theory of a self-interacting scalar field $\Phi$ 
\cite{Q},
\be
L=\partial_\mu\Phi\partial^\mu\Phi^\ast-U(|\Phi|),
\ee
where the potential satisfies the following condition,
\be
\omega^2_{\rm min}\equiv 
\min_\phi\frac{2U(\phi)}{\phi^2}<\omega^2_{\rm max}\equiv U''.
\ee
Although this 
condition can be fulfilled, for example, by a potential of the type
$U(\phi)=a\phi^6+b\phi^4+c\phi^2+d$ with $a\neq 0$, it
rules out renormalizable quartic potentials with $a=0$. As a result, the model
under consideration can at best be only some effective field theory.

\begin{figure}[t]
\begin{minipage}[c]{12cm}
  \centerline{\epsfysize=5cm\epsffile{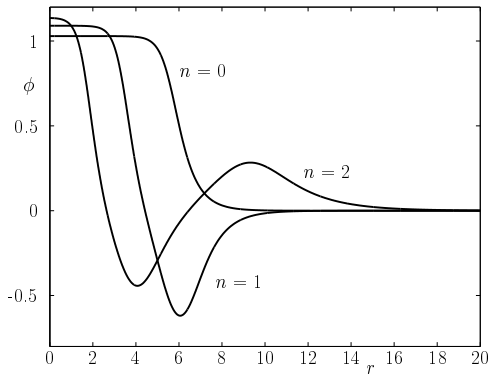}}
   \end{minipage}\hspace{1 mm}
\caption{{\fontsize{9}{9}\selectfont
The static Q-ball solutions.  
}}
\label{fig3}
\end{figure}

This theory has the rigid phase symmetry, $\Phi\to e^{i\gamma}\Phi$, 
with the associated Noether charge 
\be
Q=i\int(\dot{\Phi}^\ast\Phi-\dot{\Phi}\Phi^\ast)d^3x.
\ee
If the field does not depend on time, then Derrick's
scaling arguments apply to rule out finite energy 
solutions. If $\dot{\Phi}\neq 0$, however, then  
the existence of such solutions is not prohibited. 
These solutions, called {\sl Q-balls} \cite{Q}, are obtained
by giving to the field a time-dependent phase,
\be               \label{o}
\Phi=e^{i\omega t}\phi(r),
\ee
with real $\phi(r)$. 
The equation of motion for $\phi$ reads
\be
\phi''+\frac{2}{r}\,\phi'+\omega^2\phi=\frac{dU(\phi)}{d\phi}.
\ee
A simple qualitative analysis \cite{Q,VW02} shows that if $\omega$ is restricted 
to the range $\omega^2_{\rm min}<\omega^2<\omega^2_{\rm max}$,
then there are finite energy 
solutions for which $\phi(r)$ interpolates smoothly
between $\phi(0)\neq 0$ and $\phi(\infty)=0$. These Q-balls
form a discrete family labeled by the number 
$n=0,1,2,\ldots $ of nodes of $\phi(r)$ \cite{VW02}; see Fig.\ref{fig3}. 
Since these solutions are spherically symmetric, their energy-momentum
tensor $T^\mu_\nu$ is diagonal, and so the angular momentum 
\be
J=\int T^0_\varphi d^3x
\ee
vanishes. 

One wishes now to spin these solutions up. The idea is to give to the field a 
rotating phase \cite{VW02}, 
\be               \label{oo}
\Phi=e^{i\omega t-iN\varphi}\phi(r,\vartheta), 
 \ee
with integer $N$, where $r,\vartheta,\varphi$ are the standard spherical coordinates. 
It is worth noting that such a field is neither 
stationary nor axially symmetric. 
However, it can be called {\sl non-manifestly} stationary and 
axially symmetric for the following reason\footnote{I would like to thank 
Brandon Carter 
for clarifying discussions of this issue}.
One notices that 
the action of the generators of time translations and axial rotations 
is equivalent to the action of the rigid phase symmetry, 
\be
{\cal L}_{\partial_t}\Phi=i\omega\Phi,~~~~~
{\cal L}_{\partial_\varphi}\Phi=-iN\Phi,~~~~~
\ee 
where ${\cal L}_{\xi}$ is the Lie derivative along the vector $\xi$. 
This implies that 
the energy-momentum tensor fulfills the conditions 
\be
{\cal L}_{\partial_t}T^\mu_\nu=
{\cal L}_{\partial_\varphi}T^\mu_\nu=0,
\ee 
since it is invariant under the phase symmetry. The observable quantities 
are thus indeed 
stationary and axially symmetric. 

\begin{figure}[t]
\begin{minipage}[c]{12cm}
  \centerline{\epsfysize=5cm\epsffile{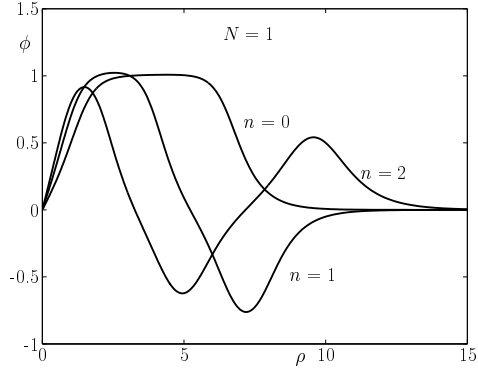}}
   \end{minipage}\hspace{1 mm}
\caption{{\fontsize{9}{9}\selectfont
The rotating Q-vortex solutions.  
}}
\label{fig4}
\end{figure}  

Before analyzing the full axially symmetric problem, it is instructive to 
consider its
simplified version in which there is the additional symmetry with respect 
to translations along the $z$-axis. The field is then given by 
\be               \label{ooo}
\Phi=e^{i\omega t-iN\varphi}\phi(\rho),
 \ee
and the field equation reduces to the ODE for $\phi(\rho)$,
\be
\left(\frac{\partial^2}{\partial \rho^2}+\frac{1}{\rho}\frac{\partial}{\partial \rho}
-
\frac{N^2}{\rho^2}+\omega^2\right)\phi=
\frac{dU(\phi)}{d\phi}.
\ee
The solutions are shown in Fig.\ref{fig4}, they describe
cylindrically symmetric configurations with a finite energy per unit length --
rotating {\sl Q-vortices} \cite{VW02}. 
Their charge $Q$ and angular momentum $J$ per unit length are related as
$J=NQ$.

Returning now to the full axially symmetric problem, 
the field equation reduces to the following PDE for $\phi(r,\vartheta)$, 
\be
\left(\frac{\partial^2}{\partial r^2}+\frac{2}{r}\frac{\partial}{\partial r}
+\frac{1}{r^2}\frac{\partial^2}{\partial \vartheta^2}
+\frac{\cot\vartheta}{r^2}\frac{\partial}{\partial \vartheta}-
\frac{N^2}{r^2\sin^2\vartheta}+\omega^2\right)\phi=
\frac{dU(\phi)}{d\phi}.
\ee
To solve this equation,  
the spectral decomposition  has been employed \cite{VW02}
\be                         \label{leg}
\phi(r,\vartheta)=\sum_{k=0}^\infty f_k(r)P^N_{N+k}(\cos\vartheta),
\ee
where $P^N_{N+k}(\cos\vartheta)$ are the associated Legendre polynomials. 
This reduces the problem to the infinite system of coupled ODE's,
\be
\left(\frac{d^2}{dr^2}+\frac{2}{r}\frac{d}{dr}-
\frac{(N+k)(N+k+1)}{r^2}+\omega^2\right)f_k(r)=
F_k(f_0,f_1,\ldots),
\ee
where $F_k$ stand for the non-linear terms. Truncating this system 
by setting $f_k=0$ for $k\geq k_{\rm max}$, gives 
a finite system of $k_{\rm max}$ coupled ODE's. 
Solutions of this truncated system have finite energy whose value 
rapidly converges to a lower non-zero limit as $k_{\rm max}$ grows. 
In fact, it turns out to be sufficient to choose $k_{\rm max}=10$
to get a reasonable approximation \cite{VW02}.

This gives spinning  $Q$-balls -- the first explicit 
example of stationary, spinning solitons
in Minkowski space in 3+1 dimensions.
For each given value of the charge $Q$, these solutions are characterized  
by the winding number $N$ and also by their parity. The latter is determined by  
whether
the index $k$ in (\ref{leg}) takes only odd or only even values.   
The distribution of the energy density is 
strongly non-spherical, and depending on the value of parity
it has the structure of deformed 
ellipsoids or dumbells oriented along the rotation axis \cite{VW02}. 
The angular momentum 
is `quantised' as 
\be
J=QN,
\ee
while the energy increases by about 20\% when the winding
number $N$ increases by one. Such a behavior
of the energy supports the interpretation of solutions with $N>0$
as describing spinning excitations of the static  Q-balls.

Perhaps the most important lesson that one can draw from this 
explicit  example of spinning 
solitons is that stationary rotation is pure field systems without gravity 
is possible. 
It is then natural  to look for spinning solitons
also in physically more interesting systems of gauge fields
with spontaneously broken symmetries.  Surprisingly, however, the results
obtained up to now in this direction are all negative. 

\section{Yang-Mills-Higgs theory}

We shall be considering a rather general class of gauge field theories
with spontaneously broken gauge symmetries \cite{VW03}. These are 
the Yang-Mills-Higgs (YMH) theories with a compact
gauge group ${\cal G}$ defined by the Lagrangian 
\be              \label{YMH}
L_{\rm YMH} = 
-\frac{1} {4} \langle F_{\mu\nu} F^{\mu\nu} \rangle
+ \frac{1} {2} \, ({\cal D}_{\mu} \Phi)^{\dagger} {\cal D}^{\mu}
\Phi - \frac{\lambda} {4} \,(\Phi^{\dagger} \Phi - 1)^2 \, .
\ee
Here, $F_{\mu\nu} \equiv {\T}_a F^a_{~\mu\nu}
= \partial_{\mu} A_{\nu} - \partial_{\nu} A_{\mu} + [A_{\mu} , A_{\nu}]$
with  $A_{\mu} \equiv {\T}_{a} A^{a}_{\mu} $. The
gauge group generators in a r-dimensional representation of ${\cal G}$ 
are ${\T}_{a}=(\T_a)_{pq}$, where  
$a = 1,2, \ldots, {\rm dim}\, {\cal G}$ and $p,q=1,2,\ldots 
{\rm r}$. They 
satisfy the relations
$[\T_{a}, \T_{b}] = f_{abc} \T_c$ and ${\rm tr}(\T_{a} \T_{b})  = K
\delta_{ab}$.
The invariant scalar product in the Lie algebra is defined as 
$\langle A B \rangle = \frac{1} {K} {\rm tr} (A B)$. The Higgs field 
$\Phi=\Phi^p$
is a vector in the representation space of ${\cal G}$ where the
generators $\T_a$ act; this space can be complex or
real. ${\cal D}_{\mu} \Phi = (\partial_{\mu} + A_{\mu}) \Phi$ is the
covariant derivative of the Higgs field. 

This field theory is quite general, and for different choices of ${\cal G}$ 
its representations it covers most of the 
known gauge models admitting solitons.  For example, 
choosing ${\cal G}$=SU(2) and the Higgs field in the adjoint 
representation, in which case the group generators are 
$(\T_a)_{ik}=-\epsilon_{aik}$, gives the theory whose solutions are 

{\bf The 't~Hooft-Polyakov monopoles} \cite{mon}. For these solutions 
the YMH fields can be chosen in the form 
\be                                     \label{3}
A^a_0=0,~~
A^a_i=\varepsilon_{aik}\frac{x^k}{r^2}(1-w(r))\,,
~~\Phi^p=\frac{x^p}{r}\,\phi(r)\,,
\ee
where indices $i,k=1,2,3$ correspond to Cartesian coordinates.
This field configuration is  spherically symmetric in the following sense. 
For each Killing vector $\xi$ generating the SO(3) rotation  group there exists
a Lie-algebra-valued scalar function 
$W_\xi$ such that the following equations are fulfilled \cite{Forgacs},
\be            \label{sym}
{\cal L}_\xi A^a_\mu=\tilde{\delta}_{W_\xi}A_\mu,~~~
{\cal L}_\xi \Phi^a=\tilde{\delta}_{W_\xi}\Phi.
\ee
Here the gauge variations induced by $W_\xi$
are defined as 
\be          \label{gauge}
\tilde{\delta}_{W_\xi} A^a_\mu=D_\mu W_\xi,~~~
\tilde{\delta}_{W_\xi}\Phi=-W_\xi\Psi,
\ee
where $D_\mu=\partial_\mu+[A_\mu,~~~]$ is the covariant
derivative in the adjoint representation. 
Conditions (\ref{sym}) 
express the invariance of the fields under the combined action
of rotations generated by $\xi$ and gauge transformations 
generated by $W_\xi$. 

Inserting (\ref{3}) to the YMH equations obtained by varying 
the Lagrangian (\ref{YMH}) 
gives a coupled system of ODE's, 
\bea
r^2w''&=&(w^2+r^2\phi^2-1)w, ~~~~\nonumber \\
(r^2\phi')'&=&2w^2\phi+\lambda r^2(\phi^2-1)\phi.
\eea
These equations admit solutions $w(r)$, $\phi(r)$
smoothly interpolating between the asymptotic
values $w(0)=1$ and $w(\infty)=0$ and $\phi(0)=0$ and $\phi(\infty)=1$. 
These are the 't Hooft-Polyakov monopoles. 
One can visualize them as extended particles containing a 
heavy core filled with the massive non-linear YMH fields, while 
only one massless component of the Yang-Mills field extends outside
the core giving rise at large distances 
to the Colombian magnetic field with unit
magnetic charge. The energy density is $O(1)$ in 
the core, while asymptotically it is $O(r^{-4})$, such that the 
total energy is finite.  

The 't Hooft-Polyakov monopoles have 
vanishing angular momentum 
-- since they are spherically symmetric. 
One can generalize these solutions to include also an electric field, which
gives electrically charged monopoles -- dyons \cite{dyon}. 
However, it is unknown 
whether one can further generalize these
solutions to include also 
an angular momentum $J\neq 0$. 
We shall now show that this is not possible {\sl within the stationary and 
axially symmetric sector}, at least for  ${\cal G}$=SU(2). 

\section{Non-existence of spinning solitons in the
Yang-Mills-Higgs theory}

Let $(A^{0}_\mu,\Phi^{0})$  be a  static, spherically symmetric soliton solution 
of the YMH theory (\ref{YMH}) for some choice 
of gauge group ${\cal G}$ and its representation. 
Let us consider 
stationary, axially symmetric on-shell deformations $(A_\mu,\Phi)$ of this solution. They 
satisfy the symmetry conditions (\ref{sym}) with 
the Killing vector $\xi$ being either stationary, 
$\xi = {\partial_t}$, or axial, 
$\xi = {\partial_{\varphi}}$. Let $W_t$ and $W_\varphi$ be
the corresponding parameters of the compensating gauge transformations 
in the right hand side of (\ref{sym}). 

The deformations are supposed to be everywhere smooth
and vanishing in the  asymptotic region,
\be
a_\mu=A_\mu-A^{0}_\mu\to 0,~~~~~
\phi=\Phi-\Phi^{0}\to 0,~~~~~{\rm as}~~~r\to\infty,
\ee
which insures that  deformed configurations belong to the same
topological sector as the original one. However, for finite values
of $r$ deformations are not supposed to be small, as long as the total energy 
of the axial configuration $(A_\mu,\Phi)$  is finite. 

To prove the absence of spinning
solitons the strategy is as follows. First, one notices \cite{VR}
that for axially symmetric fields there exists the following remarkable 
surface integral representation of the 
angular momentum:
\be
\label{JJJ}
J =\int T^0_\varphi\, d^3x 
= - \oint \langle (A_{\varphi} - W_{\varphi}) F^{k0} \rangle \ud
S_{k} \, .
\ee
Here the surface integration is performed over a closed two-surface 
expanding to spatial infinity. The calculation of the integral is then 
facilitated by 
the fact that near infinity  deformations $a_\mu$ and 
$\phi$ are small, and so they can  be described 
{\sl perturbatively}. One then carries out 
a {\sl linear}
perturbation analysis in the asymptotic region in order to decide 
whether there exist perturbations  giving a non-zero contribution
to the surface integral. 

It is worth emphasizing that results obtained in this way are 
{\sl non-perturbative}, since deviations from the static background are not 
supposed to be small everywhere. 
This allows one to draw conclusions about the existence of 
 spinning solitons for arbitrary values of $J$, without being restricted to the slow
rotation limit \cite{HSM}.

The key role in the programme outlined above certainly belongs to 
 the surface integral representation (\ref{JJJ}) of the angular 
momentum.  It is therefore important to understand where it comes from. 
It is worth noting that, normally, the angular momentum, being the Noether
charge associated with the {\sl rigid} rotational symmetry,
is given by a {\sl volume} and not surface integral. 
A surface integral representation for the Noether charge associated with a 
rigid symmetry can exist if only there is also a {\sl local}  symmetry 
in the problem, and this local symmetry contains
the rigid symmetry as a particular case. 
The Noether
current in this case has the total divergence structure typical for theories with 
local symmetries.  This implies that the volume integral for Noether's charge can 
be further transformed to a surface integral. 

A good example of such a situation can be found in General Relativity,
where asymptotic Poincar\'e symmetries in asymptotically flat 
spacetime can be viewed as a particular case of general diffeomorphisms. 
As a result, the associated Noether currents have the total divergence
structure, and the corresponding Noether charges -- mass, momentum and 
angular momentum -- are given  by the ADM {\sl surface} integrals. 

In the YMH theory under consideration the relation between
the rigid rotational symmetry and the local gauge symmetry is provided by the 
conditions
(\ref{sym}). Referring to  \cite{VW03} for details, 
the idea of how  this comes about is as follows. The Noether current
associated with the symmetry generated by a Killing vector $\xi$ 
acting on a field system with the Lagrangian 
${ L}(u^B,\partial_\mu u^B)$ is 
\be
\label{Noether}
\Theta^{\mu}_\xi  = \sum_{B} \frac{\partial {L}}
 {\partial(\partial_{\mu}u^{B})} \, \delta_\xi u^{B} - \xi^{\mu} { L} . 
\ee
In our case $u^B=(A_\mu,\Phi,\Phi^\dagger)$ and the field
variations are given by 
\be                   \label{var}
\delta_\xi A_\mu=({\cal L}_\xi-\tilde{\delta}_W) A_\mu,~~~~
\delta_\xi \Phi=({\cal L}_\xi-\tilde{\delta}_W)\Phi.
\ee
They consist of the Lie derivatives along $\xi$ and also of the pure gauge
variations generated by $W=\xi^\alpha A_\alpha$
(the gauge variations should be included to (\ref{var})  
in order to make the Noether current
gauge-invariant). Inserting (\ref{var}) to (\ref{Noether}) gives 
\be
\Theta^\mu_\xi=\xi^\nu T^\mu_\nu, 
\ee
where $T^\mu_\nu$ is the {\sl metrical} energy-momentum tensor
of the YMH system (\ref{YMH}). 
The corresponding Noether charge is given by the volume integral
\be
\Theta_\xi=\int \xi^\nu T^0_\nu d^3x. 
\ee
So far nothing new has been obtained, since this is just 
the standard expression for the conserved Noether charge associated 
with the spacetime symmetry generated by $\xi$. 
Choosing $\xi=\partial_t$
or  $\xi=\partial_\varphi$ gives the conserved energy or angular 
momentum. 

Let us now impose the symmetry conditions (\ref{sym}). 
Using these, one can eliminate 
the Lie derivatives from the variations (\ref{var}), 
which gives
\be                   \label{var1}
\delta_\xi A_\mu=(\tilde{\delta}_{W_\xi}-\tilde{\delta}_W) A_\mu
\equiv \tilde{\delta}_{\Psi_\xi}A_\mu,~~~~
\delta_\xi \Phi=(\tilde{\delta}_{W_\xi}-\tilde{\delta}_W)\Phi
\equiv \tilde{\delta}_{\Psi_\xi}\Phi,~~~~
\ee
with $\Psi_\xi=W_\xi-\xi^\alpha A_\alpha$.
As a result, the field variations for symmetric fields are 
{\sl pure gauge}
variations~! Inserting (\ref{var1}) to (\ref{Noether})  gives
\be
\Theta^\mu_\xi=-\partial_\alpha\langle\Psi_\xi F^{\alpha\mu}\rangle
-\xi^\mu L.
\ee
If $\xi=\partial_\varphi$, the second term on the right does not give 
contribution to the Noether charge,
\be                     \label{J}
J=\int \Theta^0_{\xi}d^3x=
\int\partial_k\langle\Psi_\varphi F^{k0}\rangle d^3x
=\oint\langle(A_\varphi-W_\varphi) F_{0k}\rangle dS^k.
\ee
This explains the appearance of the  surface integral representation for the  
angular momentum. 

To analyze the expression obtained, it is convenient to pass to the gauge
where the fields do not explicitly depend on $t,\varphi$, so that  
$W_t=W_\varphi=0$. The existence of such a gauge is entailed by the fact that 
the two Killing vectors $\partial_t$ and $\partial_\varphi$ commute. 
Since the deviations
$a_\mu=A_\mu-A^{0}_\mu$
from the static $J=0$ background
are small in the asymptotic region, one can linearize 
the integrand in (\ref{J}) with respect to them. This gives
\be                     \label{JJ}
J=\oint\langle a_\varphi F^{}_{0k}
+ A^{}_\varphi (-D_k a_0+[A_0,a_k ])  \rangle dS^k.
\ee
The problem therefore reduces to studying linear perturbations modes
in the asymptotic region that might give a non-zero contribution to
this integral.  

Linearizing the YMH equations around the static background $(A^{0}_\mu,\Phi^{0})$  
with respect to $a_\mu$ and $\phi$ 
gives
\bea
\label{leq1}
D^{}_{\sigma} D^{\sigma} a_{\mu}  
&-& D^{}_{\mu} D_{\sigma} a^{\sigma} +
2 [F^{}_{\mu \sigma}, a^{\sigma}] 
- {M}_{ab} \,a_{\mu}^{a}
\T_{b} \nonumber \\
&=& \frac{1} {2} \{ \phi^{\dagger} \T_{a} 
{\cal D}^{}_{\mu} \Phi^{} -
({\cal D}^{}_{\mu} \Phi)^{\dagger} \T_{a} \phi 
+ \Phi^{\dagger} \T_{a}
{\cal D}^{}_{\mu} \phi 
- ({\cal D}^{}_{\mu} \phi)^{\dagger} \T_{a} \Phi^{}
\} \T_{a} \;,  \nonumber \\
{\cal D}^{}_{\sigma} {\cal D}^{\sigma} 
\phi  &+& D^{}_{\sigma} a^{\sigma}
\Phi^{} + 2 a_{\sigma}{\cal D}^{\sigma} \Phi^{} 
 = 
-\lambda \left\{(\Phi^{\dagger} \Phi^{} - 1) \phi 
+ (\Phi^{\dagger} \phi +
\phi^{\dagger} \Phi^{}) \Phi^{} \right\} \;, \nonumber 
\eea
where the superscript `$0$' has been omitted and  
all quantities denoted by capital letters relate from now on to the 
static background. 

The long range behavior of solutions of this system is determined
by eigenvalues of the mass matrix
$
\label{mass}
{M}_{ab} = \frac{1} {2} \Phi^{\dagger} (\T_{a} \T_{b} + \T_{b}
\T_{a}) \Phi^{}$.
If all eigenvalues are positive, then all fields $a_\mu$ are massive
and tend to zero asymptotically  fast for large $r$.
The integral in (\ref{JJ}) vanishes then. 
This is the case, for example, for ${\cal G}$=SU(2) with a
doublet Higgs field, in which case the static solitons are sphalerons
\cite{sphal}. 
The conclusion therefore is that 
{the SU(2) sphalerons do not admit spinning generalizations
 within the stationary, axially symmetric sector}.  
The same conclusion can be made also in the case of vortices \cite{VW03}. 

In the case of 't Hooft-Polyakov monopoles and Julia-Zee dyons the
situation is slightly more complicated, since the mass matrix has one zero eigenvalue.
This gives rise to a long range massless component of the gauge field. 
In this case one has to solve the perturbation equations in order to determine
the asymptotic behavior  of the most
general stationary and axially symmetric perturbations $a_\mu$, $\phi$.
The corresponding general solution was found in  \cite{VW03}.
Inserting this solution to (\ref{JJ}) gives
\be
J=0
\ee
since the asymptotic inverse power-law falloff of the perturbations
turns out to be too fast to support a nonzero value of the 
integral.  This shows that {the 't Hooft-Polyakov monopoles 
and Julia-Zee dyons do not admit spinning generalizations 
within the stationary, axially symmetric sector.} 

We are therefore bound to conclude that 

\vspace{1 mm}
\noindent 
{\sl None of the known gauge field theory solitons with gauge group SU(2) --
monopoles, dyons, sphalerons, vortices --
admit spinning generalizations within the stationary, axially symmetric sector.}

\vspace{1 mm}

Of course, this does not yet eliminate completely spinning SU(2) solitons, but 
only restricts  their existence.  At the same time, this restriction is rather 
severe, since it implies that spinning counterparts for the 
known solitons, if exist at all, are not
axially symmetric. Such an option, however, seems to be rather 
implausible. The only possibility of axially symmetric, spinning solitons 
with gauge group SU(2)
that 
is still left  unexplored is related to a {\sl non-manifest} symmetry, in analogy 
with the Q-balls. 

As we have seen, in theories with a rigid phase invariance the spinning is possible
for non-manifestly axisymmetric fields 
containing rotating phases. If the invariance
is local, then the rotating phases can be gauged away, in which case the fields
are manifestly independent of $t,\varphi$.  However, 
as we have seen, the spinning is then  impossible.  At the same time, 
there exist field systems with both local and global symmetries. 
In such systems all complex fields could be given rotating phases, but
not all of these phases would be removable by local gauge transformations. 
For example, this would be the case if the dimension of a complex representation
of the gauge group is larger than the dimension of the group itself.  
In this case the number of independent gauge parameters would be insufficient
to gauge away all the  phases of the complex Higgs field. 
{\sl A non-manifestly  axially symmetric gauge field}  
would be then invariant with respect to a 
combined action of the axial symmetry plus a local gauge symmetry, 
and plus an additional  {global}
symmetry that is not a particular case of the gauge symmetry.  
A possibility of having spinning solitons in such systems remains open. 

Another possibility of constructing spinning solitons could be related to
higher gauge groups.
The pattern of the symmetry breaking and the number
of massless gauge fields that can contribute to the angular
momentum surface integral depend very much on the group. 
As a result, the possibility of having {\sl manifestly} axially symmetric rotating
solitons is not excluded for higher gauge groups.

\end{document}